# SOFTWARE ENGINEERING PRACTICE USING MULTI-CRITERIA DECISION IN HEALTH CARE ORGANIZATIONS

**Ahmed Mateen***

**Irum Tariq***

**Wajiha Azmat***

**Abstract**

A sound performance of health care organizations depends on vigorous financing system, skilled and well-paying personnel, trustworthy information, well sustained health services and reliable technologies they used. The management and organization of health care systems and other health care settings can deeply have an effect on health result, excellence of care, and patient pleasure. Environmental assessment of an organization must considered political, technical and environmental aspects. This research is paying attention on dangerous and unresolved issues resulting from waste. The insufficient managing of hospital waste be capable of creating a danger to people, in the spreading of infectious diseases, and also for the environment. Issues are classified in different groups like dangerous chemical waste, cytotoxic with carcinogenic, mutagenic and teratogenic risk and radioactive. This study focuses on the analysis of above mentioned risks and decision will be made through Multi-Criteria Decision Analysis (MCDA). MCDA techniques give considerable enhancement in decision making. Survey is used for collecting information and results are evaluated by applying some statistical tool.

*****Computer Science Department, Agriculture University Faisalabad, Pakistan**

# 1. Introduction

A good and sound performance of health care organizations and institutions depends on sound and solid financing system, fully skilled and good-paying personnel, trustworthy information, well sustained health services and reliable technologies they used. Environmental assessment of an organization must considered political, technical and environmental aspects. This research is paying attention on malignant, noxious and unresolved issues resulting from waste. Risks from above mentioned issues are classified in different groups like hazardous dangerous chemical wastage. This wastage includes toxic chemicals used in different medical equipments like cytotoxic, carcinogenic, mutagenicand teratogenic as well asradioactive. This study mainly focuses on the analysis of above mentioned risks and decision will be made through Multi-Criteria Decision Analysis (MCDA).

If wastage is categorized in this manner:

GroupA:General, no risk.

GroupB: Sanitary, assimilated to metropolitan.

GroupC: Sanitary, potentiallycommunicable.

GroupD: Bodies and human rests.

Group E: Riskychemical waste.

Group F: Cytotoxic with cancer-causing, mutagenicand teratogenicrisk.

Group G: Radioactive

MCDA (multiple criteria decision aid) has been developed as a very good source for evaluating the environment changes. It has a good capacity to control difficulties with the aim of having more objectives. These environmental difficulties have much pressure on the economy of the world. It has also effected the minds of the general public. As a result many investigations were made interrelating the economy, society and the environment. The outcome of these researches had the following characteristics:

1)      Irregularities

2)      Uncertain Probabilities

3)      Extreme dangers

4)      Disappointment

It is a much concern for the whole world. [1]

MCDA in this regard is very much helpful. It has its own type of criteria. This document initiates with a serious analysis of state-of-the-art methods for integrating various conditions in health technology assessment (HTA). The different MCDA modelling approaches are applied to a hypothetical case study. Finally, the matters that need to be measured for the use of MCDA in HTA are studied along with endorsements for future study. [2]

Health care conclusions are difficult and comprise defying trade-offs between numerous, frequently inconsistent intentions. Using organized, obvious approaches to decisions involving multiple criteria can increase the worth of decision making. MCDA approaches are broadly used in other areas, and just there has been a growth in health care applications. Health care verdicts are infrequently simple with relaxed answers. Difficulty in these verdicts is inexorable, whether a top level decision, such as that made by a financial planner, allocating limited means across actions, or at the micro level, such as a patient's decision on the top treatment substitute. Several issues influence these verdicts, a number of substitutes happen, and the information available about substitute is often unsatisfactory. [3]

Pitiable management of health care wastage possibly disclosures health care workers, wastage handlers, patients and the public at large to contamination, lethal things and harms, and dangers contaminating the environment. It is vital that all medical wastage ingredients are isolated at the point of generation, suitably cured and disposed of securely. Healthcare waste (HCW) is a creation of healthcare that comprises sharps, non-sharps, body fluid, body portions, compounds and radioactive supplies. [4]

## 2. Related Work

I.T or information technology has griped all walks of life in its hand. Similarly it has also effected the health department. With its help we can ensure the safety of medical equipments. We can also provide better and quick aid to the patients with committing no or zero mistake. The medical examination can be made online which saves the time. In short I.T has a changing effect in the field of medicine also. [5]

The present environment of healthcare facilities is scattered. However, for a good and unique way of managing the information it should be placed in a room. The facilities provided by internet helps in having control over operations that are decentralized. It improves the flow of health services. It also evokes the patients to make assessment of all the services that are being provided to them. Many systems are available on the internet that are responsible for offering health services. But these systems are unable to make privacy of the data of the patients. The privacy can only be made by a strict security system. The system should reduce all types of dangerous threats. At this point a new type of system has been developed for the treatment of hepatitis patients. For this collaboration has been made with a medicine college of kolkata. The college is main centre for HIV in India. Discussion was made with the experts from the same college to equate the characteristics that EHR was supposed to assist mainly for the Indian environment. [6]

Useof information technology assures inexpensive health facilities, helps doctors to take decisions, and assists patients to take advantages from health experts online and also enhance the efficiency of statements among healthcare suppliers and patients. The government of US gave emphasis on significance of health information technology by means of declaring every individual in state must have an electronic medical record with the hope of instant taking up health facilities. It results upgrading in health facilities and decrease in cost. Due to the development of electronic health care, medical students have given guidance during their studies duration. [7]

In underdeveloped countries several associations are running to sustain confined health care organizations and transportation to supply sustainable health care. It demands not only remedial personnel and supplies, but also needs to establish organized record of person's health care records. It also have a capability of collecting and examining data for the purpose of providing health care of citizens. A nonprofit organization developed a software application "Health records for everyone "for improving health standard in Ethiopia. By the help of this application doctor's personnel can rapidly arrange medical facilities electronically. Patient's data can also be used for field research. [8]

The basic purpose of Hospital Information System is to provide a good and sound administration and proper care to the patients with the help of digital system. This system is also responsible for providing complete medical history of all the patients. HIS is also an example of such system. The purpose of the system is to provide fast medicine services to the patients. But this system is not suitable for hospitals located in small villages as the system is expensive to build and maintain. However, this system is capable of sharing information in the same health care unit only. This system uses cloud computing system for its work. [9]

Efficient use of technology performs a necessary task in defining the influence of a field. Developed countries like Japan provides latest, timely and efficient facilities to their citizens. In understudy research a survey was conducted with the cooperation of Japan Council for Quality Health Care. Consequences of survey propose the efficiency of establishing network examination method. [10]

A general idea of chronic care modal was introduced with exacting concentration to the medical information systems, consecutively to arrange patient and residents medical information through allocation information between healthcare suppliers in organization of persistent ailments. A plan for the amalgamation of diverse services for common practitioners and hospital experts ,admittances electronic health record ,realizing the patient review and administration the swap of chronic care remedial records.[11]

As the population is increasing day by day the chronic diseases are also increasing. And its care has gained the proper attention. The present paper focuses on the chronic diseases in the China with the help of electronic health records. This record would help in preventing and treating the chronic diseases. It would also help in creating awareness and controlling to the public regarding the chronic diseases. [12]

## 3. Methodology

In health care organizations health technology assessment technologies are more considered in last year's. Other than traditional HTA techniques a new approach called" hospital-based HTA" is applied, this approach has diverse benefits and uses different methods. In health care organizations HTA techniques are concerned to tactical assessment, device administration and common managerial problems. This study sums up diverse manners to HTA performance on health organizations.

If we look at the hospitals of Pakistan we will find that plastic material is used for disposing off the waste. There is no division of this waste. Each and every thing is disposed off at the same place. The medical waste and common waste is dumped at the same place. This observation was taken with the help of research tools. It was also revealed that most of the workers who have been given the responsibility of disposing off the waste are illiterate. They are not aware of the correct way of disposing the waste. This is negligence from the side of workers. The workers also don't follow the plans that are displayed.

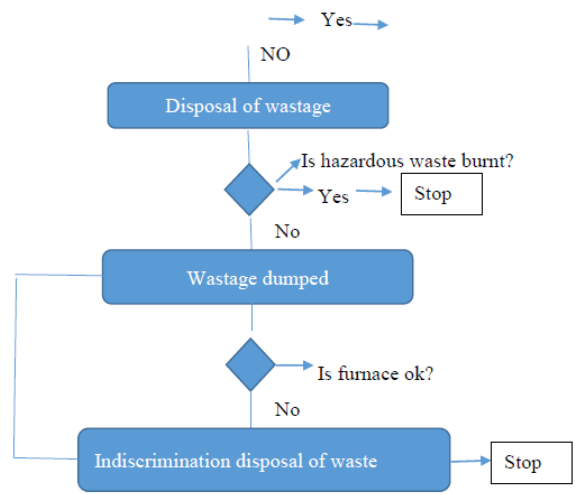

Figure 3.1.*Waste management hierarchy*

Table 3.1. Separation of health-care wastage

| Category | Type | Characteristics |
|---|---|---|
| Common | Plastic carrier | No special requirement |
| Sharps | Sealable box | Leak proof |
| Extremely dangerous | Container | Leak proof |
| Dangerous | Container | Leak proof |

The way the waste is disposed of is almost same in all the hospitals. Baskets, drums and buckets are used for disposing. The waste is usually produced by the admitted patients and their other family members who come very off and on. The waste later on is swept away by the peons who are given the responsibility of particular hospital rooms. This negligence is not prevailing at all the major and minor hospitals of Pakistan. We still have special health care centres that have fully developed staff responsible for collecting and disposing off the waste. Their team is fully aware of the proper legal and authentic way of getting rid of the waste. But the names of such hospitals are on finger tips only. Out of all the hospitals, The Cancer Hospital, Al Shifa in the capital territory, Aga Khan in the city of Quaid is worth mentioning.

One of the most prevailing and dangerous situation that is prevalent in almost all the hospitals is the combine collection of all the types of waste. Plastic made cans and boxes are usually used for this purpose. The dangerous waste is first of all collected from the ward rooms. It is then transferred to the specific tons that are made of low quality material. It is a very common observation that the waste is piled up and dumped at a place. This waste becomes a source of food for rats, cats and other small creatures which contributes in carrying the germs from one place to other which in turn brings home to a lot many diseases.

It is also noted that material that is used for making waste carriers is mostly of extreme low and poor quality that is dangerous for the health and is mostly not environment friendly. There is also no proper channel by which waste should be picked up from the hospitals. Usually it is the responsibility of the TMA. But no proper attention is given to such dumps. The dumps are either disposed of in the canals or it is burnt. The peons responsible for collecting the waste do not put on the gloves and masks so as to protect themselves for the germs. They themselves are a big source of spreading the diseases. This situation is more prevalent in public health centres as compared to the private and personally owned clinics.

The present research report shows that the most common way of getting rid of the garbage is burning in the furnace in Pakistan. It also shows that the person who is given the responsibility of the furnace is not fully aware of his responsibility and duty. There is no specification and legal advisory for the amount of the emission to be released by a particular hospital.

## 4. Results

Among the healthcare workers that participated in the study, 28.33% were medical doctors, 6.67% were pharmacists,11.67% were medical laboratory scientist,5.30% were waste handlers and 48.03% were nurses. Generally, the number of workers that have heard about healthcare waste disposal system was above average (69.5%). More health-workers in the government 81.5% than in private 57.3% hospitals were aware of healthcare waste disposalsystem and more in government hospitals attended training on it. The level of waste generated by the two hospitals differed significantly (P = 0.0086) with the generation level higher in government than private hospitals. The materials for healthcare waste disposal were significantly more available (P = 0.001) in government than private hospitals. There was no major change (P = 0.285) in syringes and needles disposal practices in the two hospitals and they were exposed to equal risks (P = 0.8510). Fifty-six (18.5%) and 140 (45.5%) of the study participants in private and government hospitals respectively were aware of the existence of healthcare waste management committee with 134 (44.4%) and 19 (6.2%) workers confirming that it did not exist in their institutions. The existence of the committee was very low in the private hospitals.

## 5. Conclusion

The investigation shows that the hospitals wastage is dangerous for health. There should be a proper plan for manage it. Environmental assessment of an organization must have considered political, technical and environmental aspects. This research is paying attention on dangerous and unresolved issues resulting from waste. The results shows the percentage of level of risk generated from hospital management, how much wastage is generated and how many healthcare organizations have a proper plan for tackle these risks. As per the study conducted in order to get rid of the wastage, burning is considered the best solution in study area. In future we plan to suggest more effective plan for managing the hospital wastage.

Table 4.1. Analysis of waste generation level in the institutions

| Types of wastes | Private % | | | Government % | | | $X^2$ | Df | P Value |
|---|---|---|---|---|---|---|---|---|---|
| | H | I | L | H | I | L | | | |
| Chemicals | 10.6 | 42.7 | 45.4 | 14.0 | 43.5 | 31.2 | 33.508 | 2 | 0.001 |
| Sharps | 44.4 | 42.7 | 12.9 | 71.1 | 16.9 | 7.1 | 72.910 | 2 | 0.001 |
| Infections waste | 25.8 | 53.3 | 20.9 | 72.1 | 20.5 | 6.8 | 1.349 | 2 | 0.001 |
| Pathological waste | 2.3 | 23.2 | 74.5 | 6.2 | 37.0 | 54.5 | 31.271 | 2 | 0.010 |
| Pharmaceuticals | 28.8 | 50.0 | 21.2 | 29.5 | 49.0 | 20.8 | 2.031 | 2 | 0.566 |
| Genotoxic waste | 5.0 | 12.9 | 82.1 | 6.2 | 23.4 | 65.3 | 31.145 | 2 | 0.001 |
| Radioactive waste | 3.4 | 18.2 | 76.8 | 5.8 | 13.6 | 74.4 | 13.176 | 2 | 0.004 |
| Domestic/General waste | 27.5 | 52.0 | 20.5 | 56.5 | 26.9 | 16.2 | 57.271 | 2 | 0.500 |

Table 4.2. Analysis on level of risks associated with different wastes

| Types of wastes | Private % | | | Government % | | | $X^2$ | Df | P Value |
|---|---|---|---|---|---|---|---|---|---|
| | H | I | L | H | I | L | | | |
| Chemicals | 21.9 | 19.5 | 7.6 | 20.1 | 16.6 | 4.9 | 4.356 | 2 | 0.225 |
| Sharps | 54.3 | 12.3 | 1.3 | 51.0 | 10.1 | 0.6 | 3.341 | 2 | 0.342 |
| Infections waste | 71.9 | 0.7 | 8.3 | 69.5 | 1.0 | 8.8 | 0.534 | 2 | 0.911 |
| Pathological waste | 9.6 | 21.2 | 2.6 | 6.5 | 23.7 | 1.9 | 2.627 | 2 | 0.453 |
| Pharmaceuticals | 7.0 | 35.4 | 21.5 | 6.5 | 35.7 | 17.2 | 2.321 | 2 | 0.508 |
| Genotoxic waste | 26.5 | 17.5 | 0.7 | 23.4 | 16.9 | 2.3 | 3.440 | 2 | 0.329 |
| Radioactive waste | 29.1 | 18.2 | 9.3 | 30.5 | 14.0 | 7.1 | 3.486 | 2 | 0.323 |
| Domestic/General waste | 0.0 | 4.0 | 21.1 | 0.0 | 1.3 | 92.2 | 4.801 | 1 | 0.091 |

H: High; I: Intermediate; L: Low

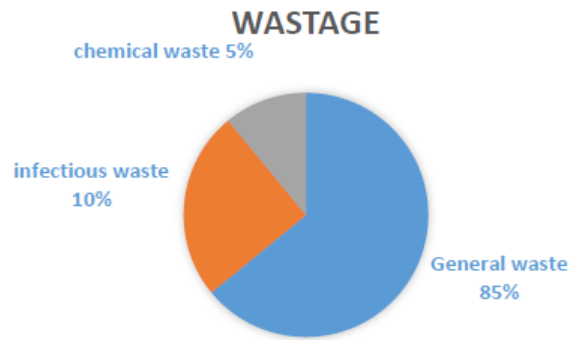

Figure 4.1. *Percentage of different types of wastages*